# Raman spectroscopic investigation of relaxor behavior in Pr doped SrTiO$_3$ and Origin of Fano resonance

Vivek Dwij, Binoy Krishna De, Shekhar Tyagi, Gaurav Sharma and V.G. Sathe*
UGC-DAE Consortium for Scientific Research,
D. A. University Campus, Khandwa Road, Indore-452001, INDIA
*Corresponding author e-mail: vasant@csr.res.in

**Abstract**

Detailed Raman spectroscopy studies on polycrystalline Sr$_{1-x}$Pr$_x$TiO$_3$ (x=0.01, 0.025, 0.05, 0.075, 0.09, 0.13, 0.15, 0.17) samples are reported elucidating the microscopic mechanism of relaxor ferroelectrics. The polar mode was observed upto very high temperature ~1000 K suggesting that the dipoles exists at temperatures well above the characteristic relaxor temperatures and they develop a short range correlation at 505 K leading to formation of PNRs. The TO$_2$ polar mode showed anomalous softening in cooling below T~505 K supporting the growth of PNRs. Fano resonance is reported in the Pr doped compounds which decreases with increasing doping. Our work on Pr doped SrTiO$_3$ reveals that the local TiO$_6$ octahedral tilt scales with the intensity of the polar mode and hence can be used as an order parameter for relaxor transition. The study showed that the paraelectric to relaxor ferroelectric phase transition in this compound is random polarizability instability driven phenomenon which is correlated with the local octahedral tilt angle. The study supports competition and cancelation between lattice and polar instabilities at global length scale while cooperation between the two at local length scale. The modulation in local structure of the material in temperature interval related to dielectric anomaly has been observed which is not been investigated previously by any structural tool.

**Introduction:**

Relaxor ferroelectrics are important technological materials and hence been investigated actively for long time due to their superiority over normal ferroelectrics [1,2,3,4]. The difference between the ferroelectrics and relaxor ferroelectrics comes from the correlation length of the off-centred ions. Another noticeable property of relaxors is huge but diffuse dielectric maxima it displays when measured as a function of temperature ($T_m$) and dispersion in $T_m$ when measured as a function of frequency. These properties are generally attributed to the growth and dynamics of the polar nano regions (PNRs) in the matrix of a paraelectric phase. The Polar Regions are produced by off-centering of A-site and/or B-site ions in the perovskite systems, such as displacement of Pb ions from its high symmetry position along with displacement of B-site cations which is Nb in case of Pb(Zn$_{1/3}$Nb$_{2/3}$)O$_3$ [5,6,7,8,9] producing non-centrosymmetry [10,11] at local length scale. The local off-centered regions and induced relaxor behaviour comes from the local fields generated due to compositional fluctuation of the doped ions [12,13]. The nucleation of PNRs is marked by a temperature known as Burns temperature $T_B$. However, it is still not clear as to how the local structural polarity develops with temperature in relaxors. Below $T_B$ several characteristic temperatures marking the properties of PNRs are defined in literature that are summarised by Mihailova et al [14] as follows: (i) a fully dynamic stage above $T_B$, (ii) quasidynamic behavior between $T_B$ and $T^*$; $T^*$ is defined as the temperature below $T_B$ where initial freezing of the polar nano regions start, (iii) quasistatic between $T^*$ and $T_f$: $T_f$ is defined as the freezing temperature observed just below $T_m$ where the system undergoes from ergodic to non-ergodic state with static coupled PNRs and (iv) static below $T_f$. Thus in a relaxor ferroelectrics the length scale and dynamics of PNRs play a major role in shaping the physical response of the system. Though $T_B$ and $T_f$ are fairly well understood, the microscopic origin of $T^*$ is still not well established [15,16]. Further, a recent theoretical study showed existence of Fano resonance in relaxor ferroelectrics occurring due to interference of the phonon frequencies corresponding to two different atoms in a doped compound [17]. In the same study it is proposed that the T* defines a temperature below which the polar mode frequencies related to Ti-sublattice hardens significantly.

Disorder is the key to relaxor behavior and it is responsible for modulation of the local structure in relaxor materials. Disorder can arise from different routes such as chemical substitution, defect, strain etc. Strontium Titanate ($SrTiO_3$) being a quantum paraelectric is highly susceptible to disorder making it incipient ferroelectric system. Strain, defects, isotope substitutions, doping, external electric field makes a polar ground state of the material. Interestingly, such polar state is directly reflected in the Raman spectrum as polar modes which are a direct signature of the polarity and hence it should provide direct information about the growth, dynamics and quasi-static freezing of the PNRs denoted by the characteristic temperatures of the relaxor system as mentioned above. It is quite fascinating to note that profile of these Raman mode differs under different local environment. The profile of the polar $TO_2$ mode has been related with the local polar correlations and shown to represent autocorrelation function of fluctuating polar nano-regions (PNRs) [18,19]. However, the profile of the polar modes in $SrTiO_3$ were not well understood. Although a lot of work is reported about the detection of the PNRs by x-ray diffraction, neutron diffraction, polarized Raman spectroscopy and IR spectroscopy [3,4,20,21,22,23], there are very few reports on the effect of growth and dynamics of the PNRs on the polar modes [24,25].

In order to investigate the behaviour of polar modes in relaxors, we chose Pr doped $SrTiO_3$ system. $SrTiO_3$ (STO) is a known quatum paraelectric which shows tetragonal transition through antiferro distortion (AFD) below 105 K. This compound, thus shows AFD modes below 105 K in Raman spectroscopy but the spectra remains devoid of polar modes. When doped with Pr which has small ionic radii than the Sr ions, the system shows ferroelectricity at room temperature [26]. This compound showed centrosymmetric cubic structure for low $x$ concentration ($x \leq 0.075$) and tetragonal structure at high doping values. The ferroelectricity in this compound was highly debated because of the presence of the global centrosymmetric structure. The issue is settled when Garg et al [27] conclusively showed from their high resolution x-ray and neutron diffraction measurements that the system is globally centrosymmetric but retains non-centrosymmetric distortions at local level and the behaviour is best described in the frame work of relaxor ferroelectrics. Checchia *et al* [28] carried out detailed x-ray diffraction studies on this series of compounds and observed room-temperature tetragonal structure (*I4/mcm*) for compositions with $x \geq 0.1$, whereas cubic structure for compositions with $x \leq 0.075$ at room temperature. The existence of PNRs at room temperature was confirmed by the observation of polar modes in the Raman spectroscopy study in all the doped compositions at room temperature. It was also shown that the local tilting angle of the oxygen octahedra surrounding the Ti ions is active in all the compositions independent of their long range symmetry. Their experimental results also supported coexisting AFD and FE instabilities predicted by Aschauer and Spaldin [29] in their theoretical studies.

In this report we carried out detailed Raman spectroscopy studies from 80 K to 1018 K on polycrystalline $Sr_{1-x}Pr_xTiO_3$ (x=0.01, 0.025, 0.05, 0.075, 0.09, 0.13, 0.15, and 0.17) samples. We believe that in compounds with x≤0.075 the smaller Pr ion can act similar to defects inducing disorder while in compounds with x>0.075 the Pr incorporation in higher concentration should induce chemical pressure. In all the samples the $TO_2$ polar mode showed anomalous behaviour upto temperature which is between $T_m$ and $T_B$ and hence attributed to $T*$ and the polar mode showed waning with increasing temperature but could be observed up to 1018 K in all the samples. We also observed asymmetric profile of the Polar modes establishing presence of Fano resonance.

## Experimental:

We have used SrTiO₃ single crystal for $x = 0.00$ composition in the present study which is a commercially brought substrate (MTI Corporation, USA) and the rest of the $Sr_{1-x}Pr_xTiO_3$ ceramic compounds ($x$= 0.010, 0.025, 0.050, 0.075, 0.09, 0.13, 0.15, and 0.17) were prepared using conventional solid state reaction route. Stoichiometric ratios of the high purity (99.99%) $SrCO_3$, $Pr_6O_{11}$, $TiO_2$ (Alpha Aesar) precursors were mixed using a mortar and pestle for 6-8 hours with acetone as an intermediate media. The mixed samples were calcined at 1150°C for 10 hours. After cooling, the calcined sample was grinded again for 6-8 hours and pressed axially in a 10 mm pallets. These pallets were sintered at 1420°C for 12 hours. The process was repeated multiple times to obtain the main phase. The obtained samples were characterized by high resolution X-ray diffraction and Raman spectroscopic measurements. High resolution X-ray diffraction measurements were performed using Cu Kα radiation Lab source (on both of Bruker and Rigaku diffractometer) with step size of 0.02°. The Raman measurements were carried out using Horiba JY HR-800 spectrometer with an 1800 g/mm grating and a CCD detector. We used a He-Ne excitation source (633 nm) laser beam focused into ~1μm diameter spot in backscattering geometry where the incident light is linearly polarized and spectral detection is unpolarized. An Olympus microscope was used to view the images of the surface of the sample and an LMplanFI 50X objective lens for focusing. The overall spectral resolution of the system is ~1 cm⁻¹. The temperature dependent Raman measurements were performed by mounting the pressed pellets in Linkam THMS600 stage for the low temperature measurements (80-300 K) and in Linkam TS1000 stage for the high temperature measurements (300-1018 K), having a temperature stability of ±0.1 K. The collected Raman spectrums has been corrected using Bose-Einstein occupation factor as $I = I_{collected}/\{1 + (e^{\hbar\omega/k_BT} - 1)^{-1}\}$, where $\hbar,\omega,k_B,T$ are reduced Plank's constant, phonon wavenumber, Boltzmann's constant and Temperature respectively. All the Raman spectrum presented are normalized with respect to SrTiO₃ second order mode.

## Results

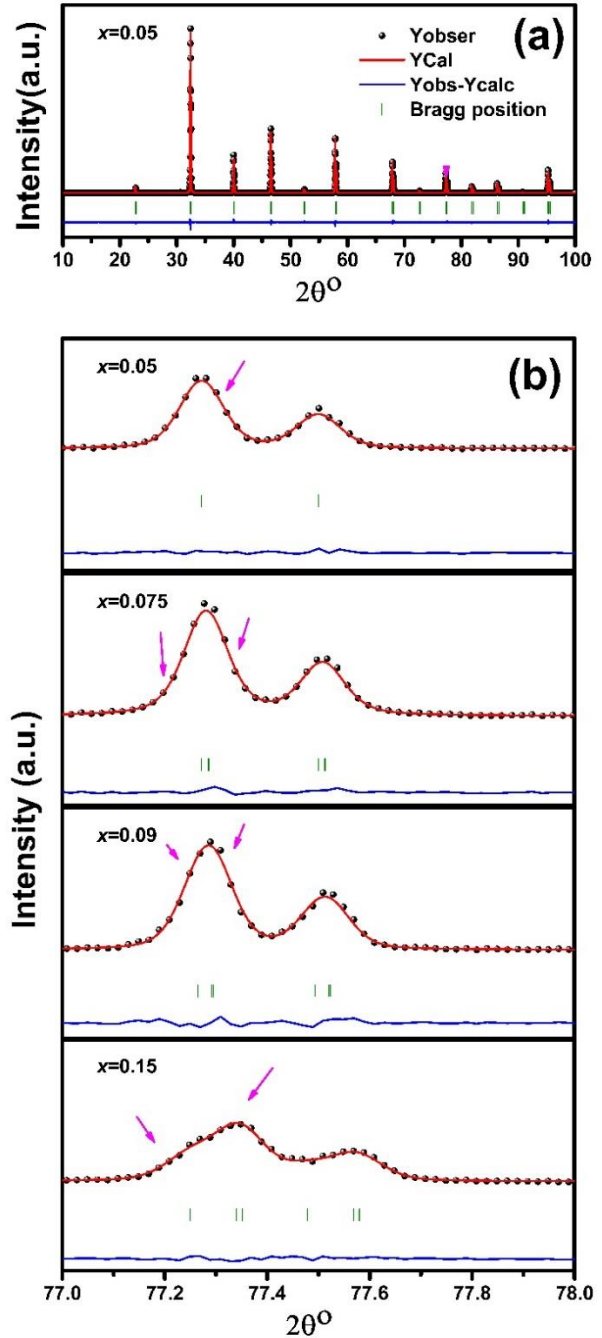

**Figure 1:** (a) The experimental powder x-ray diffraction pattern for $x$=0.05 (dots) along with fitted curve obtained using Rietveld refinement (line) and the difference plot at the bottom of the curve. (b) The enhanced view of the x-ray diffraction pattern depicting asymmetry and splitting of the peaks. The vertical ticks in the figure represents the position of the Bragg reflections.

The x-ray diffraction measurements were carried out in order to assess the long range symmetry of the samples and FullProf software [30] was used to refine the diffraction patterns. A representative diffraction pattern for x=0.05 along with fitting and difference curve is shown in figure 1(a). The patterns could be well fitted by using cubic symmetry with $Pm\bar{3}m$ space group for compositions $x<0.075$ with satisfactory goodness of fit parameters ($\chi^2 \sim 1.6$). Garg et al [27] reported tetragonal symmetry for $x \geq 0.05$ compositions from neutron and Raman scattering measurements while Checchia et al [28] reported cubic symmetry up to $x=0.075$ from their x-ray diffraction measurements. The $x \geq 0.05$ composition was therefore attempted to be fitted with cubic as well as tetragonal space group (I4/mcm), however, the $x=0.05$ was found to be fitting better with cubic symmetry. Figure 1 (b) showed enhanced view of the diffraction pattern between 2θ=77-78°. It showed asymmetry in the peaks for $x \geq 0.075$ compositions which increases with increasing $x$. It is a signature of tetragonal symmetry and the corresponding splitting in this peak for $x \geq 0.075$ increases with increasing $x$. A careful observation of figure 1 (b) clearly shows the increase in asymmetry due to separation of d values corresponding to (116), (332) and (420) reflections with increase in tetragonal distortion. In perfect cubic symmetry all the three peaks fall at the same $d$ value.

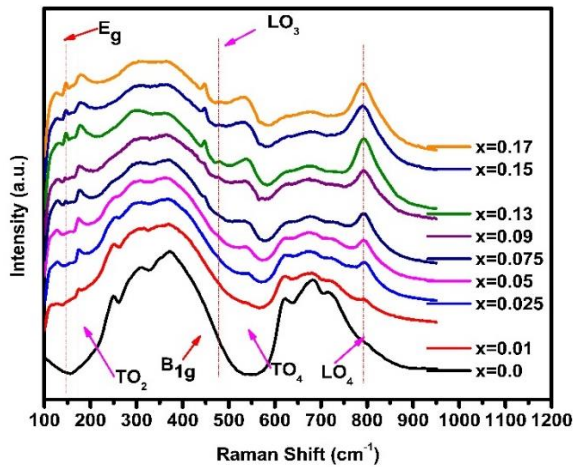

**Figure 2:** The Raman spectra of $Sr_{1-x}Pr_xTiO_3$ ($x=$ 0.0-0.17) collected at room temperature.

In order to access the local symmetry of these samples, room temperature Raman characterization was carried out and is presented in Figure 2. For low doped compositions i.e. $x=0.01$ and 0.025 five Raman modes are observed at 117, 127, 175, 540, 795 cm$^{-1}$; the position of these modes matched with previous reports [28,31,32,33]. As such in SrTiO$_3$; first order Raman scattering is forbidden in the cubic symmetry shown by this compound at room temperature, but the presence of guest or impurity atoms (Pr in present case) relaxes the selection rules and gives rise to first order scattering. Apart from these modes a feeble mode is observed at 475 cm$^{-1}$ which is assigned as LO$_3$ mode in the literature. For $x \geq 0.075$ composition, two additional modes are observed at 145 cm$^{-1}$ and 445 cm$^{-1}$. The position of these modes matches with $E_g$ and $B_{1g}$ modes observed in pure SrTiO$_3$ below 105 K. These modes arise out of AFD in tetragonal symmetry due to bending of octahedron as a$^o$a$^o$c$^-$ [28] following Glazer scheme of notations. From previous results, [28, 3231] we have assigned the remaining Raman modes falling at 175, 540, 795 cm$^{-1}$ as TO$_2$, TO$_4$ and LO$_4$ modes, respectively. The TO$_2$ mode corresponds to polar last mode representing motion of Sr/Pr with respect to TiO$_6$ octahedron while TO$_4$ mode corresponds to polar axe mode arising due to bending of the octahedron [31]. Presence of the TO$_2$ and TO$_4$ modes in the Raman spectra has been considered as an evidence of presence of the polar instability [18,34] in the system. The distortion of the unit cell from cubic to tetragonal at lower temperature originates from antiferrodistortive (AFD) transition which makes the $E_g$ and $B_{1g}$ modes Raman active. $E_g$ and $B_{1g}$ modes represent activation of TiO$_6$ octahedral tilt around c-axes which occurs from condensation of triply degenerate R$_{25}$ phonon [28,32]. It is observed that with increase in $x$, the intensity of the TO$_2$ and TO$_4$ polar modes increases. The line width of the TO$_4$ mode also showed an increase with increasing composition. Also the position of the TO$_2$ mode shows a shift towards higher wavenumber as the doping is increased. The increase in wavenumber with increasing $x$ can be attributed to the hardening of the force constant of the lattice vibration due to incorporation of heavier Pr atoms in place of Sr atoms [35]. The increased line width with increasing $x$ suggests decrease in phonon life time due incorporation of the larger site disorder in the system. Another important observation of the room temperature results is the presence of intense TO mode. Such strong intensity of the TO modes were not reported in other compounds. This can be due to preserving nature of Pr ions towards ferroic character [36] which not only causes first order scattering due to local symmetry lowering but also these distortions are strongly coupled to optical modes and hence enhances local scattering [37]. Further, the $E_g$ mode also shows a hardening with doping which is related to the octahedral tilt. Finally, we observed asymmetric profile of $TO_2$ mode which becomes nearly symmetric for $x\geq0.09$. Such asymmetric profile is observed in doped $SrTiO_3$ and $KTaO_3$ at low temperatures and is attributed to Fano resonance [18,19,34]. Interestingly, polar $TO_2$ and $TO_4$ Raman modes coexist with AFD induced $E_g$ amd $B_{1g}$ modes in this material for $x\geq 0.075$. This is against the general understanding of the competition among the antiphase octahedral rotations and polar displacement [38,29] resulting in absence of polar modes in the undoped $SrTiO_3$. From XRD and Raman characterization, we thus conclude that globally the studied compositions show cubic symmetry for $x<0.075$ while compositions with $x\geq0.075$ shows signatures of tetragonal symmetry at room temperature while the local symmetry of all the doped samples is lower than the cubic symmetry. $LO_4$ mode observed in Raman spectra which arises due to introduction of Pr in STO matrix becomes broad and intense with increasing doping concentration indicating higher disorder in the system with higher doping.

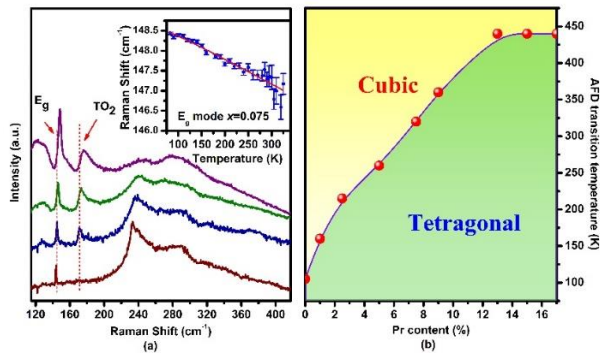

**Figure 3:** (a) The Raman spectra of $Sr_{1-x}Pr_xTiO_3$ ($x=$ 0.0, 0.010, 0.025, 0.075) collected at 83 K (bottom to top). The inset shows the variation of $E_g$ mode intensity as a function of temperature for $x$=0.075. (b) The AFD transition temperature as a function of $x$.

The asymmetric profile of the polar $TO_2$ mode and coexisting polar and AFD distortions needs more clarity. For the same low temperature Raman spectra were collected for low doped compositions. Figure 3 (a) shows the normalized Raman spectra collected on these compositions at 83 K. The spectra are shifted vertically for clarity. For better comparison, the Raman spectra of the $SrTiO_3$ single crystal is also included. It is noted that the $TO_2$ mode ($\sim$174 cm$^{-1}$) is observed in all the compositions except for undoped $SrTiO_3$. This mode showed hardening, increase in intensity and decrease in asymmetry with increasing $x$. The line width also showed significant increase with increasing $x$. Similarly, the $E_g$ mode which is a signature of tetragonal symmetry (AFD) is observed in all the samples and its intensity, line width and position showed an increase with increasing $x$. The increase in line width of the $E_g$ mode reflects the site disorder induced due to Pr doping. It is worth noting that the intensity of polar mode ($TO_2$) as well as AFD mode ($E_g$) show an increase with increasing $x$. This clearly suggests the one to one correspondence between polarities of the sample and the tetragonal distortion. As mentioned before this is against the general paradigm that the polar distortion and AFD competes and cancels [29,41]. From this it can be concluded that the AFD and Polar distortions both increases with increasing $x$ at local length scale. The behavior of $E_g$ mode position as a function of temperature for $x$=0.075 composition is plotted in the inset of Figure 3 (a). It showed a normal behavior that is nearly linear decrease in intensity and softening with increase in temperature and it gets completely suppressed above 320 K. It is noted from this figure that the AFD distortions increases with increasing $x$. We have monitored the tetragonal distortion as a function of temperature and $x$ by tracking the AFD ($E_g$ and $B_{1g}$) modes. Based on the disappearance of these modes, we could deduce the AFD transition temperature. AFD transition increases linearly with doping upto x=0.09, while for x>0.09 its value is nearly constant. Also, asymmetry in polar $TO_2$ mode is maintained till the lowest temperature indicating absence of any additional Raman mode besides the $TO_2$ mode. To obtain information from polar $TO_2$ mode we attempted different methods including multiple Lorentzian as well as Fano function for fitting. For Fano function fitting, we have used the function $I(\omega) = I_c * \frac{|q+\varepsilon|^2}{1+\varepsilon^2} + I_b$ where $\varepsilon = \frac{\omega - \omega_p}{\Gamma}$, q is the asymmetry parameter, $\omega_p$ is renormalized phonon frequency, and $\Gamma$ is the line width [39]. It was observed that the $TO_2$ mode can be better fitted using Fano function at all temperature in comparison to the multiple or single Lorentzian. The $TO_2$ mode for various $x$ fitted using the Fano function is shown in figure 4. It can be seen that peak position and width of the Raman mode increases linearly while asymmetry parameter showed a nonlinear behavior Figure 4(b); it initially shows a linear increase before showing an abrupt jump in the tetragonal phase.

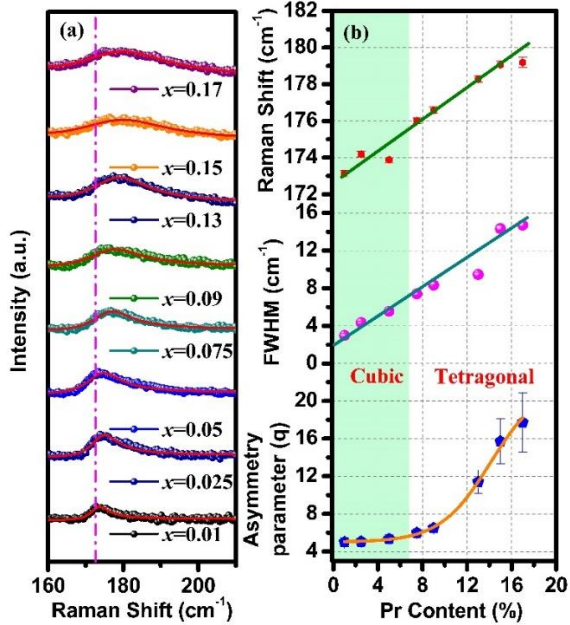

**Figure 4:** (a) $TO_2$ mode for various compositions along with fitted curve using Fano function. (b) Raman shift, width and asymmetry parameter resulted from fitting is plotted as a function of $x$. (Green) Shaded region represents cubic symmetry while un-shaded region represents tetragonally distorted structure.

Since Fano asymmetry is retained for x<0.09 compositions, we carried out temperature dependent Raman spectroscopy studies from room temperature (300K) to 1018 K so as to observe the changes in the Raman spectra across the $T_m$, $T^*$ and $T_B$. The Raman spectra collected at elevated temperatures on ($x$=0.01, 0.025, 0.05 and 0.075) showed expected changes with increasing temperature i.e. the Raman modes showed suppression and increased line width with increasing temperature. Most of the modes diminished at higher temperatures except the $TO_2$ mode. The spectra collected at 1018 K is presented in figure 5. At such an elevated temperature also, $TO_2$ mode is observable for these doped compositions. This suggests that the weak local polar distortions remain present in all the doped compositions up to very high temperatures.

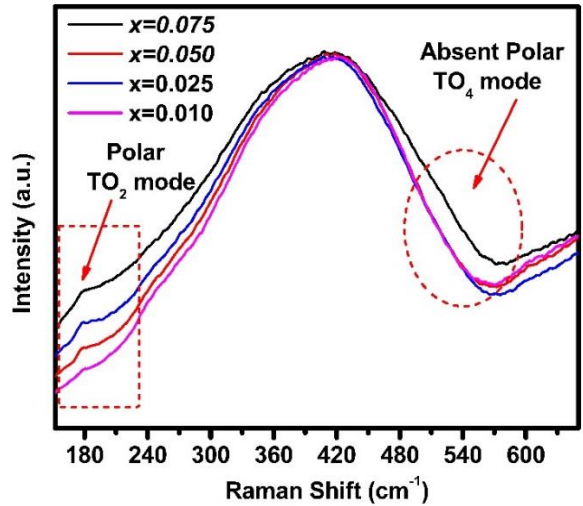

**Figure 5:** Raman spectrum at 1018K for various compositions.

The $TO_2$ mode is fitted using Fano function and the position and asymmetry of the $TO_2$ mode was obtained as a function of temperature. The mode position as well as asymmetry parameter of the $TO_2$ mode as a function of temperature for the four compositions are plotted in figure 6. It showed anomalous behavior of the mode wavenumber up to a temperature which is in between $T_m$ and $T_B$, the wavenumber showed an increase with increasing temperature. After this the temperature variation of the mode frequency showed a change in slope and then it shows a constant variation at elevated temperatures, up to 1018 K. The mode wavenumber of the TO$_4$ mode was observed to show monotonous normal behavior (not shown here). For Fano fitting function integrated intensity is proportional to $I_c|q|^2$ [19]. The integrated intensity of the x=0.01 and 0.025 composition normalized with respect to the intensity of the second order mode of the SrTiO$_3$ has been plotted in figure 7. It showed a distinct change in slope at temperature which is again in between T$_m$ and T$_B$. Checchia et al showed one to one correspondence between the tilt angle of the oxygen octahedra and intensity of the TO2 modes in this system [28]. Therefore, dramatic variation in intensity below 505 K in this samples suggests changes in oxygen octahedral tilt angle.

dielectric anomaly region while the vertical dotted (pink) line marks the T*=505K.

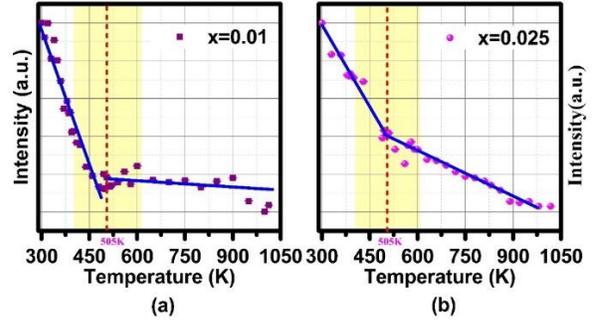

**Figure 7:** TO$_2$ mode intensity variation as a function of temperature for x=0.01 and x=0.025.

### Discussion:

The present studies showed three interesting phenomenon which are interrelated with each other. The first is the observation of anomalous softening of TO$_2$ mode below 505 K under cooling. The second is the variation in octahedral tilt angle which showed a nearly linear increase below 505 K under cooling and the third is the vanishing of asymmetric line shape of polar mode above 505 K. In order to find out the correlation among doping and octahedral tilt, we calculated octahedral tilt angle for various doping from our room temperature XRD measurements. It may be noted here that the E$_g$ mode originates from TiO$_6$ octahedral tilt around c-axes and hence the E$_g$ mode position or wavenumber as a function of doping percentage of Pr ions should show a strong correlation with octahedral tilt. The normalized tilt angle obtained from XRD analysis and shift in E$_g$ mode position are plotted as a function of Pr concentration in figure 8(a). Interestingly it showed a one to one correlation establishing strong correlation of Eg mode position with octahedral tilt angle. In fact, the temperature dependent variation in the shift of Eg followed mean field equation $\phi(T) = \phi(0)[1 - \frac{T}{T_c}]^\beta$ where β=0.5 (see figure 8(b) in agreement with the second order nature of this transition as reported previously [28]. It is worth noting here that not only the polar mode position and octahedral tilt showed

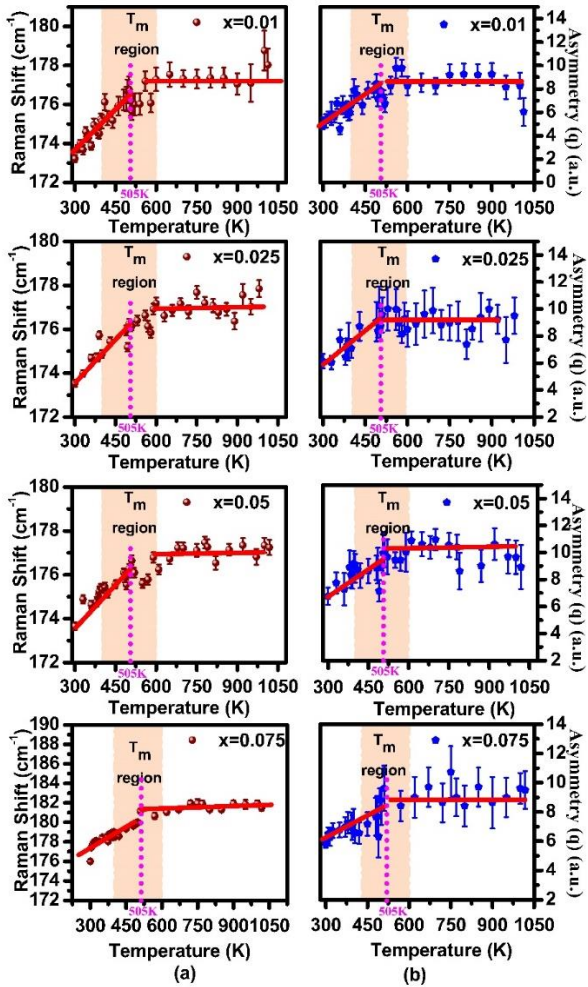

**Figure 6:** Temperature dependence of TO$_2$ mode position and Asymmetry parameter, Here the data is represented by dots while the red line is a guide to the eye. The shaded region (orange) marks the strong correlations but the normalized asymmetry profile shape parameter "$q$" deduced by fitting the TO2 mode using Fano line shape showed one to one correspondence (see figure 8(a)).

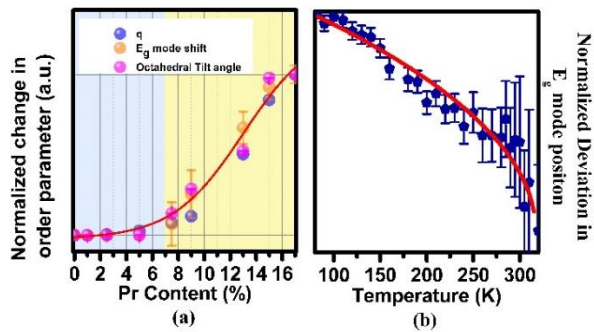

**Figure 8:** (a) Normalized variation in order parameters (asymmetry parameter, $E_g$ mode shift and octahedral tilt angle as a function of Pr concentration. (b) Temperature dependence of normalized deviation in Eg mode frequency.

The asymmetry in the TO2 polar mode can be thought of arising due to small sizes of the PNRs, however it is well known that the size of the PNRs decreases with increasing temperature while the asymmetry (~1/q) [1818,19] showed an decrease with increasing temperature ruling out this possibility. The other possibility for asymmetric profile arises due to Fano resonance. Fano resonance arises from interference of electronic continuum state with discrete phonon state as observed in case of $CaCu_3Ti_4O_{12}$ and $SrCu_3Ti_4O_{12}$ [39]. However, here the sample is highly insulating and there are no reports showing local metallic states. Recent theoretical calculations predicted strong Fano resonance in $BaZr_{0.5}Ti_{0.5}O_3$ relaxor ferroelectric [17]. In this report, it is stated that the phonons from Zr and Ti sublattices interferes with each other causing Fano resonance and the motions of the Ti and Zr ions become coupled to each other as a result of this Fano coupling. The calculations suggested that the Ti-oscillator is strongly damped providing the continuum state which interacts with the sharp discrete phonon state provided by the Zr-oscillator resulting in Fano resonance. In present system, the PNRs are formed due to A-site chemical disorder which leads to local regions with a distribution of fluctuating/ dynamical off-centered Ti positions along with regions with non-fluctuating off-centered Ti ions. Therefore, the continuum state in Pr doped $SrTiO_3$ can be a response of dynamic disorder in Ti atom position i.e. rapidly fluctuating local polar correlations originating from A-site induced disorder. The interference of this continuum state with discrete $TO_2$ polar mode is attributed for the Fano resonance.

The Asymmetry in profile of the $TO_2$ mode showed a decrease with increasing x. This asymmetric profile is quantified numerically with asymmetry parameter ($q$) given as [40]

$$q = \frac{\langle \Phi|T|i \rangle}{\pi V_E^* \langle \Psi_E|T|i \rangle}$$

Where $\Phi$ is renormalized state due to mixing of continuum state $\Psi_E\rangle$ and discrete state $|\phi\rangle$ with correlation interaction ($V_E^*$) with transition matrix T and ground state $|i\rangle$. The observed value of $q$ depends on the renormalized state of the system due to interference between the two. In our system, the polar continuum and the intensity of the $TO_2$ mode are highly correlated [18, 19]. Therefore, with increasing temperature both the polar continuum diminishes with decreasing $I_{TO2}$. This would decrease the $\langle \Psi_E|T|i \rangle$ term which in turn result in increase in $q$ value. Also at elevated temperatures coupling among discrete and continuum state is expected to decrease and hence $V_E^*$ should also decrease. The asymmetry reflects presence of fluctuating polar correlations. The $q$ value represents the correlation between polar fluctuations and polar phonon. Lower the value of $q$ parameter larger the correlation as $q$ is inversely proportional to configuration interaction in the system. With increasing $x$, value of $q$ showed an increase suggesting lower correlation between polar order and phonon at higher Pr concentration. As mentioned before as the $x$ is increased the long range AFD instability appears resulting in tetragonal structure. The long range AFD order hinders the local polar order or fluctuation in polar order. This explains the loss of asymmetry or increasing $q$ value with increasing $x$. It is worth mentioning here that the width of the $TO_2$ mode showed an increase with increasing $x$ which is a signature of competition and cancelation of the polar order with setting in of long range AFD.

The observation of polar modes up to very high temperatures (~1000K) well above the Tm, T* and TB in all the compositions suggest that microscopically the dipole exists even at such elevated temperatures. This is against the present understanding that the dipoles form around Burns temperature. It seems that the local dipoles form well above the characteristic temperatures and they order locally or at short range around a temperature generally attributed to T* which is near the Tm. The anomalous softening of the polar mode below 505 K or below T* indicates towards renormalization of the phonon frequencies due to setting in of the short range dipolar order. It is likely that the setting in of short range order modulates the lattice whereby renormalizing the associated phonon mode frequency. In present case the $TO_2$ polar mode showed a direct correlation with setting in of short range order. In such situation one expects a change in the octahedral tilt around T* as the octahedral tilt is responsible for the development of the polar order. As mentioned before the intensity of the $TO_2$ mode is directly related with the local octahedral tilt. Figure 8 shows that the intensity of this mode changes its slope across T*. This shows that the octahedral tilt is getting modulated due to short range polar order. It may be worth mentioning here that we have already established the cooperation among local AFD instabilities and local polar order [41]. The one to one correspondence between octahedral tilt angle, shift in $E_g$ mode wavenumber and asymmetry parameter depicting Fano resonance again supports the above paradigm. The anomalous softening of $TO_2$ polar modes is also reported in PbTiO3 above the ferroelectric transition temperature and is attributed to the local ordering of A-site and B-site induced dipoles [42].

Present understanding associates T* with the coupling of initially nucleated small polar clusters and their aggregation into larger polar clusters [14]. The temperature ranges between $T_B$ and T* is characterized by a coupling between adjacent off-centered $BO_6$ octahedra to form initial polar clusters, while the range between T* and $T_m$ is characterized by a coupling between off-centered B cations from adjacent polar clusters. Our observation of first order Raman scattering at elevated temperatures ($>>T_B$) suggests that $BO_6$ octahedral remains locally deformed. Local octahedral tilt is the major structural distortion in the system. This distortion causes two Ti-sublattices in the Pr doped system. These sublattices remain incoherent at elevated temperatures ($>T_B$). The coupling among these two Ti-sublattices results in local structural modifications (octahedral tilt here) generating a polarization component and increase in polar cluster size (q decreases as $I_{TO2}$ increases). In analogy to classical ferroelectrics, polarization extension occurs due to flattening of multiwall potential hence the respective temperature should mark local phase transition as intermediate temperature (T*) ~ 505K. This induced polarization component renormalizes the polar phonon frequency as a spectroscopic signature of T* and the dynamic coupling between these two sublattices manifests in the form of Fano resonance in the system.

### Conclusions: -

This study thus clearly brought out the microscopic mechanism involved in relaxor ferroelectrics. It is shown that microscopic polar distortions exist well above the characteristic temperatures. This suggests that the relaxor ferroelectric behavior is a random polarization instability driven phenomenon and around the characteristic temperatures like $T_B$,

T* and Tm short range correlations among the polar distortions or dipoles appear leading to formation of PNRs. The formation of short range correlations modulates the lattice leading to variation in octahedral tilt and renormalization of polar phonon mode frequencies. The fluctuating Ti off-centered ions interferes with discrete polar modes leading to asymmetric line shape due to Fano resonance. The AFD and polar instabilities competes and cancels at global length scale while they cooperate at local length scale. These results are in consonance with microscopic picture for ferroelectric paraelectric phase transition in $PbTiO_3$ where it is described as a stochastic polarization instability driven by a progressive misalignment instead of a complete disappearance of the local dipoles [42].